\documentclass[12pt]{amsart}
\usepackage{amsmath, amssymb, latexsym, amsthm}
\usepackage[dvips]{graphicx}

\newcommand{\mx}[1]{\left (\begin{matrix}#1\end{matrix}\right )}

\newcommand{\RND}{\mathsf{RND}}
\newcommand{\PFF}{{\mathcal{P}_{\mathrm{FF}}}}
\newcommand{\F}{\mathcal{F}}
\newcommand{\abs}[1]{{\left |#1\right |}}
\renewcommand{\O}{\mathcal{O}}
\newcommand{\A}{\mathcal{A}}
\newcommand{\B}{\mathcal{B}}
\newcommand{\C}{\mathcal{C}}

\renewcommand{\(}{\left (}
\renewcommand{\)}{\right )}

\newcommand{\cyc}{\op{cyc}}
\newcommand{\comp}{\op{comp}}
\newcommand{\x}{\times}
\newcommand{\Bad}{\mathsf{Bad}}

\newcommand{\bbR}{\mathbb{R}}
\newcommand{\bbZ}{\mathbb{Z}}
\newcommand{\Om}{\Omega}
\newcommand{\be}{\begin{enumerate}}
\newcommand{\ee}{\end{enumerate}}
\newcommand{\bi}{\begin{itemize}}
\newcommand{\ei}{\end{itemize}}
\renewcommand{\i}{\item}
\newcommand{\sm}{\setminus}
\newcommand{\op}[1]{\operatorname{#1}}
\newcommand{\OCS}{\op{OCS}}

\newcommand{\Prob}{\op{Pr}}
\newcommand{\f}[2]{\frac{#1}{#2}}
\renewcommand{\P}{\mathcal{P}}
\newcommand{\inv}{^{-1}}
\long\def\forget#1\forgotten{}

\newtheorem{thm}{Theorem}[section]
\newtheorem{prop}[thm]{Proposition}

\newtheorem{lem}[thm]{Lemma}
\newtheorem{cor}[thm]{Corollary}
\newtheorem{conj}{Conjecture}

\theoremstyle{definition}
\newtheorem{definition}[thm]{Definition}
\newtheorem{example}[thm]{Example}

\theoremstyle{remark}
\newtheorem{rem}[thm]{Remark}

\newcommand{\Part}[1]{\part{#1}}

\title[Fast forward permutations]{Permutation graphs, fast forward permutations, and
sampling the cycle structure of a permutation}
\author{Boaz Tsaban}
\address{Department of Mathematics and Computer Science, Bar-Ilan University,
Ramat-Gan 52900, Israel}
\email{tsaban@macs.biu.ac.il, http://www.cs.biu.ac.il/\~{}tsaban}

\begin{document}
\begin{abstract}
$P\in S_N$ is a \emph{fast forward permutation} if for each $m$
the computational complexity of evaluating $P^m(x)$ is small
independently of $m$ and $x$.
Naor and Reingold constructed fast
forward pseudorandom cycluses and involutions. By studying the
evolution of permutation graphs, we prove that the number of
queries needed to distinguish a random cyclus from a random
permutation in $S_N$ is $\Theta(N)$ if one does not use queries of
the form $P^m(x)$, but is only $\Theta(1)$ if one is allowed to
make such queries.

We construct fast forward permutations
which are indistinguishable from random
permutations even when queries of the form
$P^m(x)$ are allowed.
This is done by introducing an efficient
method to sample the cycle structure of a
random permutation,
which in turn solves an open problem of Naor and Reingold.
\end{abstract}

\maketitle

\setcounter{section}{-1}
\section{Introduction and Motivation}

According to Naor and Reingold \cite{NaRe},
a permutation $\sigma\in S_N$ is a
\emph{fast forward permutation} if for each
integer $m$, and each $x=0,\dots,N-1$, the computational complexity of
evaluating $\sigma^m(x)$ is small and independent of
$m$ and $x$. An important example for such a permutation is
the \emph{successor} permutation $s$ defined by
$$s(x) = x+1 \bmod N,$$
as for each $m$ and $x$,
$s^m(x) = x+m \bmod N$.
Observe that $s$ is a \emph{cyclus}, that is, its cycle
structure consists of a single cycle of length $N$.

Throughout this paper, the term \emph{random} is
taken with respect to the uniform distribution.
In \cite{NaRe}, Naor and Reingold consider the following problem%
\footnote{
For the sake of clarity, we will concentrate in the beginning in the
(purely) random case, and leave the \emph{pseudo}random
case for Part \ref{PRcase}.
}:
Assume that we have a fast forward permutation
$\sigma\in S_N$.
Assume further we have an oracle%
\footnote{
An \emph{oracle} is an algorithm initialized by a fixed
unknown initial state, which works as a ``black box''
by accepting queries of some specific form,
and making responses accordingly.
(The initial state of the algorithm may change as it runs.)
The user of such an algorithm can only know the queries and the responses to them.
}
$\P$ which fixes
a random permutation $P\in S_N$, and for each
$x$ can compute $P(x)$ and $P\inv(x)$ in
time which is polynomial in $\log N$.
We wish to use this oracle in order to define
a random permutation $Q$ such that:
\be
\i $Q$ is a random element of the space of all permutations which have
the \emph{same cycle structure} as $\sigma$.
\i $Q$ is a \emph{fast forward permutation}.
\ee
The solution to this problem is as follows \cite{NaRe}:
Define $Q=P\sigma P\inv$. Then for each integer
$m$ we have that
$$Q^m(x) = P(\sigma^m(P\inv(x))),$$
so $Q$ is a fast forward permutation.
Moreover, $Q$ has the same cycle structure
as $\sigma$, and it is not difficult to
see that it distributes uniformly among the
permutations which have the same
cycle structure as $\sigma$.

Therefore Naor and Reingold's construction using $\sigma=s$
yields a fast forward random cyclus.
The natural question which arises is whether this construction
gives a pseudorandom \emph{permutation}.
Here by \emph{pseudorandom permutation} we mean that the
resulting permutation is difficult to distinguish from a truly
random permutation using a limited number (under some reasonable
definition of ``limited'') of calls to the oracle.
In Section 4 of \cite{NaRe} it
is conjectured that distinguishing a random cyclus in $S_N$ from
a random permutation should require roughly $\sqrt N$ evaluations.
In the forthcoming Section \ref{indist} we prove
that in the restricted model where
only queries of the form $P(x)$ or $P\inv(x)$ are allowed
(this is the usual model),
the task of distinguishing a random cyclus from a random permutation
requires roughly $N$ (not $\sqrt N$) evaluations.

However, if one wants to allow the usage of the fast
forward property
in the mentioned construction then the resulting permutation is
far from being pseudorandom: In Section \ref{crypt}
we show that a single evaluation is enough to
distinguish a random cyclus from a random permutation
in the fast forward model (where
evaluations of the form $P^m(x)$ are allowed).
Therefore, the question of construction of a
fast forward pseudorandom permutation is far
from having a satisfactory solution.
It turns out that a solution of this problem
can be obtained by solving another open problem.

After introducing their construction, Naor and Reingold
ask whether it is possible to remove the restriction
on the cycle structure of the fast forward permutation,
that is, whether one can use the oracle $\P$
in order to define a random permutation $Q$ such that:
\be
\i $Q$ is a random element in the
space $S_N$ of \emph{all} permutations.
\i $Q$ is a \emph{fast forward permutation}.
\ee
We give an affirmative solution which is based on
an efficient method to sample the cycle
structure of a random permutation,
together with an introduction of
a fast forward permutation for
any given cycle structure.
This construction yields a
fast forward random permutation which is
indistinguishable from a random permutation
even in the fast forward model.


\Part{Indistinguishability and distinguishability}
This part deals with the
evolution of permutation graphs and its application to
the indistinguishability of random cycluses from random permutations,
and with the distinguishability of random cycles from
random permutations when fast forward queries are allowed.

\section{The indistinguishability of random cycluses from random permutations}
\label{indist}

In this section we prove that the number of evaluations of the form
$P(x)$ or $P\inv(x)$
needed in order to distinguish a random cyclus in $S_N$ from
a random permutation in $S_N$ is $\Theta(N)$.

Our proof is best stated in the language of graphs.
We first set up the basic notation and facts.
As these are fairly natural, the reader may wish to
skip directly to Lemma \ref{firstlemma},
and return to the definitions
only if an ambiguity occurs.

Throughout this section, $V=\{0,\dots,N-1\}$
and $G$ (with or without an index) will denote
a finite directed graph
with $V$ as its set of vertices.

Fix a natural number $N$. The graph of a (partial) function $f$
from (a subset of) $N$ to $N$ is the directed graph with set of
vertices $V$ and with an edge from $x$ to $y$ if, and only if,
$f(x)=y$ (for all $x,y\in V$). For convenience we also require
that for all $x,y\in V$ there exists at most one edge from $x$ to
$y$, and will write $x\to y$ when there exists an edge from $x$
to $y$. The graph of a (partial) function will be called a
\emph{(partial) function graph}. Observe that there is a natural
bijective correspondence between (partial) functions and their
graph. A particular case of (partial) function graphs is the
(partial) permutation graph, where we require that the (partial)
function of the graph is injective.

Let $\Phi$ denote the ``forgetful'' functor assigning to each
directed graph $G$ the corresponding undirected graph $\Phi(G)$
(each edge from $x$ to $y$ is replaced by an undirected edge
between $x$ and $y$.)
A set $C$ of vertices in $G$ is a \emph{component}
if it is a connected component in the undirected
graph $\Phi(G)$ (isolated vertices are also components).
A component $C$ is \emph{connected}
if for each $x,y\in C$ there exists a path from $x$ to $y$
in $G$.

If $G$ is a partial function graph
then each connected component of $G$ is a \emph{cycle}.
A permutation graph $G$ of a cyclus will be called a
\emph{cyclus graph}. Thus a cyclus graph
has a single connected component, and has the form
$$x_0\to x_1\to\dots\to x_{N-1}\to x_0.$$
$G$ is a \emph{partial cyclus graph} if it can be extended
to a cyclus graph. A partial cyclus graph is \emph{proper} if it is
not a cyclus graph.

The following sequence of observations
will play a key role in our proof.
We will give proofs only where it seems necessary.

\begin{lem}\label{firstlemma}
Let $G$ be a directed graph.
The following are equivalent:
\be
\i $G$ is a proper partial cyclus graph.
\i $G$ is a partial permutation graph with no cycles.
\i Each component of $G$ is well-ordered by $\to$.
\ee
\end{lem}

Thus if $G$ is a proper partial cyclus graph then
each component $C$ of $G$ contains a unique
minimal element $\min C$ and a unique
maximal element $\max C$.

\begin{lem}\label{cycextension}
Assume that $G$ is a partial cyclus graph with $m$ components.
Then there exist
exactly $(m-1)!$ cyclus graphs extending $G$.
\end{lem}
\begin{proof}
Let $C_0,\dots,C_{m-1}$ be the components of $G$.

Fix any \emph{cyclus} $\sigma\in S_m$. For each $i=0,\dots,m-1$,
add an edge from $\max C_{\sigma^i(0)}$ to $\min C_{\sigma^{i+1}(0)}$
to obtain a cyclus graph $G^\sigma$.
We claim that for distinct cycluses $\sigma,\tau\in S_m$,
the graphs $G^\sigma$ and $G^\tau$ are distinct.
Indeed, let $i\in\{0,\dots,m-1\}$ be the minimal such that
$\sigma^{i+1}(0)\neq\tau^{i+1}(0)$ (observe that
$\sigma^0(0)=0=\tau^0(0)$.)
Then in $G^\sigma$ there is an edge from
$\max C_{\sigma^i(0)}$ to $\min C_{\sigma^{i+1}(0)}$, whereas in
$G^\tau$ there is not. Thus each cyclus in $S_m$ defines a unique
cyclus graph extending $G$.

On the other hand, each cyclus graph extending $G$ defines a unique well-ordering
on $G$ by removing the edge pointing to $\min C_0$, and this well-ordering
defines, in turn, a unique cyclus $\sigma\in S_m$ by letting
$\sigma^{i+1}(0)$ be the unique $k$ such that there is an edge from
$\max C_{\sigma^i(0)}$ to $\min C_{k}$.

It remains to recall that there exist exactly $(m-1)!$ cycluses in $S_m$.
\end{proof}

Let $\comp(G)$ and $\cyc(G)$ denote the collection of components and
cycles in $G$, respectively.
The following lemma describes the basic steps in the evolution of
partial permutation graphs. We use $\uplus$ to denote disjoint union.

\begin{lem}\label{evolution}
Assume that $G$ is a partial permutation graph, and
let $\tilde G$ be the new graph obtained by adding
a new edge to $G$. Then $\tilde G$ is a partial permutation graph if,
and only if, there
exist (not necessarily distinct) connected components $C_0$ and
$C_1$ in $G$ such that
the new edge is from $\max C_0$ to $\min C_1$. Moreover,
\be
\i If $C_0$ and $C_1$ are the same component then
$\comp(\tilde G) = \comp(G)$, and
$\cyc(\tilde G) = \cyc(G)\uplus\{C_0\}$.
(In particular, $|\comp(\tilde G)| = |\comp(G)|$, and
$|\cyc(\tilde G)|=|\cyc(G)|+1$.)
\i If $C_0$ and $C_1$ are distinct then
$\cyc(\tilde G)=\cyc(G)$, and
$\comp(\tilde G) = (\comp(G)\sm\{C_0,C_1\})\uplus\{C_0\cup C_1\}$.
(In particular,
$|\cyc(\tilde G)|=|\cyc(G)|$, and
$|\comp(\tilde G)|=|\comp(G)|-1$.)
\ee
\end{lem}


For the following definition,
recall our convention that throughout this paper, the term \emph{random} is
taken with respect to the uniform distribution.
\begin{definition}\label{oracle12}
Define the following oracles:
\bi
\i[$\C$:] Chooses a random cyclus $P\in S_N$,
accepts queries of the form $(x,i)\in \{0,\dots,N-1\}\x \{1,-1\}$
and responds with $y=P^i(x)$ for each such query.
\i[$\O_2$:] Begins with the empty graph $G_0$ on $V=\{0,\dots,N-1\}$,
accepts queries of the form $(x,i)\in V\x \{1,-1\}$,
and constructs a partial cyclus graph on $V$ as follows.
In the $k$th query $(x_k,i_k)$,
the oracle responds as follows:
\be
\i If the query was made earlier and answered with $y$,
or a query of the form $(y,-i_k)$ was made earlier and answered
with $x_k$, then the oracle responds with $y_k=y$.
\i Otherwise, the oracle responds as follows (let $C_{x_k}$ denote
the component of $x_k$):
\be
\i If $i=1$ then it chooses a random $C\in\comp(G_k)\sm\{C_{x_k}\}$,
sets $y_k = \min C$,
adds the edge $x_k \to y_k$ to $G_k$ to obtain
a new graph $G_{k+1}$, and responds with $y_k$.
\i If $i=-1$ (this is the dual case)
then it chooses a random $C\in\comp(G_k)\sm\{C_{x_k}\}$,
sets $y_k = \max C$,
adds the edge $y_k \to x_k$ to $G_k$ to obtain
a new graph $G_{k+1}$, and responds with $y_k$.
\ee
\ee
\ei
A sequence $((x_0,i_0),y_0,\dots(x_k,i_k),y_k)$
is \emph{$\C$-consistent} if the equations $P^{i_j}(x_j)=y_j$
have a solution $P\in S_N$ which is a cyclus.
It is \emph{nonrepeating} if there exists no
$0\le j<l\le k$ such that $(x_l,i_l)=(x_j,i_j)$, or
$(x_l,i_l)=(y_j,-i_j)$. Thus a nonrepeating sequence
is a sequence where Case 1 of $\O_2$ is
never activated, that is, a sequence in which each
query answer gives new information on the permutation (or its
graph). Observe that any consistent sequence can be turned
into a shorter nonrepeating sequence which induces the
same partial cyclus graph.
\end{definition}

\begin{lem}\label{sameproc}
For each nonrepeating $\C$-consistent sequence\\
$s=((x_0,i_0),y_0,\dots(x_{k-1},i_{k-1}),y_{k-1})$,
$$\Prob[s | \C] = (N-k-1)!/(N-1)! = \Prob[s | \O_2],$$
where $\Prob[s | \A]$ is the probability that the oracle $\A$
responds with $y_0$ to $(x_0,i_0)$, then with $y_1$ to $(x_1,i_1)$, \dots,
and finally with $y_{k-1}$ to $(x_{k-1},i_{k-1})$.
\end{lem}
\begin{proof}
The definition of $\C$-consistency ensures that the
sequence $s$ defines a partial cyclus graph.
The requirement that $s$ is nonrepeating implies
by Lemma \ref{evolution}
that each answer to a query
reduces the number of components in the induced
partial cyclus graph by exactly $1$.
Thus, after $k$ queries the induced graph has
exactly $N-k$ components. By Lemma \ref{cycextension},
there exist $(N-k-1)!$ cyclus graphs extending the
given partial cyclus graph, and therefore the probability
of getting $s$ in $\C$ is $(N-k-1)!/(N-1)!$.

Now consider $\O_2$.
Again, Lemma \ref{evolution} implies that
$|\comp(G_j)|= N-j$ for all $j$. Given $G_j$, the probability for
a specific consistent answer $y_j$ in the next query to $\O_2$ is
$1/(N-j-1)$ (uniform choice of one out of the remaining
$N-j-1$ components).
Thus,
$$\Prob[s | \O_2] = \f{1}{N-1}\cdot\f{1}{N-2}\cdot\ldots\cdot\f{1}{N-k} = \f{(N-k-1)!}{(N-1)!}.$$
\end{proof}

We say that two oracles are \emph{equivalent} if there is no way to distinguish
between them by making queries to the oracles and analyzing their responses.
\begin{cor}\label{cycequiv}
The oracles $\C$ and $\O_2$ are equivalent.
\end{cor}

\begin{definition}\label{oracle3}
Define the following oracles.
\bi
\i[$\O_3$:] Initially sets a flag $\Bad$ to $0$,
and begins with the empty graph $G_0$ on $V=\{0,\dots,N-1\}$.
This oracle accepts queries of the form $(x,i)\in V\x \{1,-1\}$,
and constructs a partial \emph{permutation} graph on $V$ as follows.
In the $k$th query $(x_k,i_k)$,
the oracle responds as follows:
\be
\i If the query was made earlier and answered with $y$,
or a query of the form $(y,-i_k)$ was made earlier and answered
with $x_k$, then the oracle responds with $y_k=y$.
\i Otherwise, the oracle responds as follows:
\be
\i If $i=1$ then it chooses a random $C\in\comp(G_k)$, sets $y_k = \min C$,
adds the edge $x_k \to y_k$ to $G_k$ to obtain
a new graph $G_{k+1}$, and responds with $y_k$.
\i If $i=-1$ (this is the dual case)
then it chooses a random $C\in\comp(G_k)$, sets $y_k = \max C$,
adds the edge $y_k \to x_k$ to $G_k$ to obtain
a new graph $G_{k+1}$, and responds with $y_k$.
\ee
If $C$ is the component of $x_k$, this oracle sets
$\Bad = 1$.
\ee
\i[$\P$:] Chooses a random permutation $P\in S_N$,
accepts queries of the form $(x,i)\in \{0,\dots,N-1\}\x \{1,-1\}$
and responds with $y=P^i(x)$ for each such query.
\ei
\end{definition}

A sequence $((x_0,i_0),y_0,\dots(x_k,i_k),y_k)$
is \emph{$\P$-consistent} if the equations $P^{i_j}(x_j)=y_j$
have a solution $P\in S_N$.
The proof of the following is similar to the proof
of Lemma \ref{sameproc}
(in fact, it is simpler) and we omit it.
\begin{lem}
For each nonrepeating $\P$-consistent sequence $s$
which corresponds to $k$ queries and replies,
$$\Prob[s | \O_3] = (N-k)!/N! = \Prob[s | \P].$$
\end{lem}

\begin{cor}\label{permequiv}
Oracles $\O_3$ and $\P$ are equivalent.
\end{cor}

For our purposes it seems convenient to use the following notion of
a distinguisher.
An (information theoretic) \emph{distinguisher} $D$ is a probabilistic algorithm%
\footnote{
A \emph{probabilistic algorithm} is an algorithm enhanced by an access to a random
number generator, that is, at each stage the algorithm chooses which moves to make next
according to some well-defined distribution. Mathematically, a probabilistic algorithm
is a random variable, whereas a usual algorithm is a function.
}
with
an unlimited computational power and storage space, which accepts an oracle as input (where there are
two possible oracles),
makes $m$ queries (where $m$ is some fixed number)
to that oracle (the distribution of each
query depends only on the sequence of earlier queries and oracle responses),
and outputs either $0$ or $1$
(again, the distribution of the answer depends only on the sequence
of queries and oracle responses).

The intended meaning is that the distinguisher's output is its guess as
to which of the two possible oracles made the responses.
(Thus given two oracles $\A$ and $\B$, $D(\A)$ and $D(\B)$ are
random variables taking values in $\{0,1\}$.)
The natural measure for the effectiveness of the distinguisher
in distinguishing between
two oracles $\A$ and $\B$ is its
\emph{advantage}, defined by
$$|\Prob[D(\A)=1]-\Prob[D(\B)=1]|.$$
The motivation for this measure is as follows.
Assume without loss of generality that
$\Prob[D(\A)=1]\ge\Prob[D(\B)=1]$.
Then by the likelihood test
we should decide $x=\A$ if the output of $D(x)$
is $1$ and $x=\B$ otherwise. The effectiveness
of this decision procedure clearly increases as
the difference between
$\Prob[D(\A)=1]$ and $\Prob[D(\B)=1]$ increases,
and this (or any other) procedure is useless when the probabilities
are equal. Moreover, it can be proved that the number
of times needed to sample $D(x)$ in order to decide whether
$x=\mathcal{A}$ or $x=\mathcal{B}$ with a significant level of certainty is
$O(1/\epsilon^2)$, where $\epsilon = |\Prob[D(\A)=1]-\Prob[D(\B)=1]|$.

\begin{thm}\label{cycvsperm}
Assume that $D$ is a distinguisher which makes $m<N$ queries to
$\C$ or $\P$.
Then
$$|\Prob[D(\C)=1]-\Prob[D(\P)=1]|\le \f{m}{N}.$$
\end{thm}
\begin{proof}
By Corollaries \ref{cycequiv} and \ref{permequiv},
it suffices to show that $|\Prob[D(\O_2)=1]-\Prob[D(\O_3)=1]|\le \f{m}{N}$.

Oracles $\O_2$ and $\O_3$ behave identically as long as $\Bad = 0$ in $\O_3$, that is,
as long as the component of $x_k$ was not chosen.
As long as this is the case, the number of components in the graph reduces by at most $1$ with
each new query answer (we do not assume that the queries are nonrepeating), and therefore
the probability that the component of $x_k$ was not chosen for all $k=0,\dots,m-1$ is at least
$$\f{N-1}{N}\cdot\f{N-2}{N-1}\cdot\ldots\cdot\f{N-m}{N-m+1} = \f{N-m}{N} = 1-\f{m}{N}.$$
Let $p = \Prob[D(\O_2)=1]$. Then
$p = \Prob[D(\O_3)=1|\Bad=0]$, therefore
\begin{eqnarray*}
\lefteqn{\Prob[D(\O_3)=1] =}\\
& = & \lefteqn{\Prob[D(\O_3)=1|\Bad=0]\cdot\Prob[\Bad=0] +}\\
& & + \Prob[D(\O_3)=1|\Bad=1]\cdot\Prob[\Bad=1]\\
& = & p\cdot\Prob[\Bad=0] + \Prob[D(\O_3)=1|\Bad=1]\cdot\Prob[\Bad=1].
\end{eqnarray*}
Thus,
\begin{eqnarray*}
\lefteqn{|\Prob[D(\O_2)=1]-\Prob[D(\O_3)=1]| =}\\
& = & |p(1-\Prob[\Bad=0]) - \Prob[D(\O_3)=1 | \Bad=1]\cdot\Prob[\Bad=1]|\\
& = & |p\cdot\Prob[\Bad=1]-\Prob[D(\O_3)=1 | \Bad=1]\cdot\Prob[\Bad=1]|\\
& = & |(p-\Prob[D(\O_3)=1 | \Bad=1])\cdot\Prob[\Bad=1]|\le\\
& = & |p-\Prob[D(\O_3)=1 | \Bad=1]|\cdot\f{m}{N}\le \f{m}{N}.
\end{eqnarray*}
\end{proof}

\begin{cor}
For all $\epsilon>0$, the
number of evaluations required to distinguish a random cyclus in $S_N$ from a
random permutation in $S_N$ with advantage greater or equal to $\epsilon$
is at least $\lfloor \epsilon N\rfloor$.
\end{cor}

Our bound on the distinguisher's advantage cannot be improved.
The following theorem shows not only that there exists an
optimal strategy (with advantage $m/N$) for the distinguisher,
but that in some sense \emph{all} strategies are optimal,
including for example those which do not use queries of the form $(x,-1)$.
By ``all'' we mean those which do not make queries where
the responses are known in advance, that is, strategies
for which the sequence of queries is nonrepeating.
(As we remarked before, any strategy which makes repeating
queries can be improved.)

\begin{thm}[Optimal strategies]\label{optimal}
Consider the following $m$-step strategy ($m<N$) for a distinguisher $D$ to distinguish between
$\P$ and $\C$:
\bi
\i[Queries:] For each $k=0,\dots,m-1$, choose any pair
$(x_k,i_k)\in V\x \{1,-1\}$
such that the sequence $((x_0,i_0),y_0,\dots,(x_k,i_k))$ is nonrepeating,
and make the query $(x_k,i_k)$.
\i[Output:] If one of the oracle responses introduced a cycle,
the distinguisher outputs $1$. Otherwise the distinguisher outputs $0$.
\ei
Then the advantage of this distinguisher is $m/N$.
In other words, any strategy which generates only
nonrepeating sequences is optimal.
\end{thm}
\begin{proof}
As the query sequence is nonrepeating, the probability that a cycle is not introduced
given that the oracle is $\O_3$ is
exactly
$$\f{N-1}{N}\cdot\f{N-2}{N-1}\cdot\ldots\cdot\f{N-m}{N-m+1} = \f{N-m}{N} = 1-\f{m}{N}.$$
Thus $\Prob[D(\P) = 0] = \Prob[D(\O_3) = 0] = 1-m/N$, and
$$\Prob[D(\C) = 0]-\Prob[D(\P) = 0] = 1-\(1-\f{m}{N}\)=\f{m}{N}.$$
\end{proof}

\section{Cryptanalysis of the Naor-Reingold fast forward cyclus}
\label{crypt}

In this section we show that in the fast forward model
(where the distinguisher is allowed to make queries of the form $P^m(x)$),
random cycluses can be distinguished from random permutations
with advantage $1-o(1)$, using a single query to
the given oracle.

For each $N$ let $d(N)$ denote the number of divisors of $N$.
\begin{thm}\label{checkcyclus}
A fast forward random cyclus can be distinguished from
a fast forward random permutation with advantage $1-d(N)/N$,
using a single query.
\end{thm}
\begin{proof}
We will use the following important fact.
\begin{lem}[folklore]\label{uniformcyclen}
Fix an $x\in\{0,\dots,N-1\}$. Then the length of the cycle of $x$
in a random permutation in $S_N$ distributes uniformly in
$\{1,\dots,N\}$.
\end{lem}
\begin{proof}
For each $k=1,\dots,N$ the probability that the cycle's length is $k$ is
$$\f{N-1}{N}\cdot\f{N-2}{N-1}\cdot\ldots\cdot\f{N-(k-1)}{N-(k-2)}
\cdot\f{1}{N-(k-1)}=\f{1}{N}.$$
\end{proof}
Assume that $P$ is a random permutation in $S_N$. By Lemma \ref{uniformcyclen},
the length $a_0$ of the cycle of $0$ distributes uniformly in
$\{1,\dots,N\}$. As there are $d(N)$ divisors of $N$,
the probability that
$a_0$ divides $N$ is $d(N)/N$.
Now, $P^N(0)=0$ if, and only if, $a_0$ divides $N$.
Thus, the probability that $P^N(0)=0$ is $d(N)/N$
if $P$ is random, but $1$ if $P$ is a cyclus.
Therefore, the single query $(0,N)$ is enough to distinguish a
random cyclus from a random permutation with advantage $1-d(N)/N$.
\end{proof}

\begin{example}
If $N=2^n$ (this is the standard case), then $d(N)/N=(n+1)/2^n$,
which is negligible.
\end{example}

$d(N)/N$ converges to $0$ quite rapidly as $N\to\infty$.
However, for our purposes, the following easy observation is
enough.
\begin{prop}
$d(N)/N=o(1)$.
\end{prop}
\begin{proof}
Observe that for each $N$, if the factorization of $N$ is
$p_1^{e_1}\cdot\ldots\cdot p_k^{e_k}$, then
$d(N)=(e_1+1)\cdot\ldots\cdot(e_k+1)$, thus
$$\f{d(N)}{N} = \f{e_1+1}{p_1^{e_1}}\cdot\ldots\cdot\f{e_k+1}{p_k^{e_k}}.$$
For all $N>1$, as the function $f(x)=(x+1)/N^x$ is decreasing for
$x\ge 0$, we have that for all $k\ge 1$, $(k+1)/N^k\le 2/N\le 1$.

Fix any $\epsilon>0$.
If $N$ has a prime factor $p\ge 2/\epsilon$, then
$d(N)/N\le 2/p\le\epsilon$.
Otherwise, all prime factors of $N$ are smaller than $c=2/\epsilon$.
Assume that $N=p_1^{e_1}\cdot\ldots\cdot p_k^{e_k}$.
Then $k\le c$. Let $e_i=\max\{e_1,\dots,e_k\}$.
$N\le c^{e_1+\dots+e_k}$, so $ce_i\ge e_1+\dots+e_k\ge\log_c N$,
therefore $e_i\ge h(N)=\log_c N/c$, thus
$d(N)/N\le (e_i+1)/p_i^{e_i}\le (h(N)+1)/p_i^{h(N)}$ which is
smaller than $\epsilon$ for large enough $N$.
\end{proof}

\begin{rem}\label{rem1}
One may suggest the following ad-hoc solution to the
problem raised by Theorem \ref{checkcyclus}:
Simply bound the possible value of $m$ in queries of the
form $P^m(x)$ to be $\le N/k$ for some fixed $k$.
But then $P^N(x)$ can still be computed (using $k$ queries
instead of $1$), so this solution is not good if
we do not want to restrict the value of $m$ too much.
\end{rem}

\begin{rem}\label{generalization}
Theorem \ref{checkcyclus} can be extended as follows: Fix a cycle
structure. Let $a_0$ be the size of the largest cyclus in this
structure, and assume that $P\in S_N$ is a random permutation with
the given cycle structure. The probability that an element $x$
appears in a cyclus of size $a_0$ is (at least) $a_0/N$. If $k$ is
$\Om(N/a_0)$, then with large probability one of the elements
$0,\dots,k-1$ appears in the cyclus and therefore $P^{a_0}(i)=i$
for some $i\in\{0,\dots,k-1\}$. But if $P$ is random, then it is
conceivable that with a non-negligible probability (it is not
straightforward to quantify the term ``non-negligible'' here), for
all $i\in\{0,\dots,k-1\}$ the cycle lengths do not divide $a_0$
and therefore $P^{a_0}(i)\neq i$.

Of course, if $a_0<N/a_0$, then one may simply verify in $a_0$
calls that the cycle of $0$ has size $\le a_0$. Thus
our method works in complexity $O(\min\{a_0,N/a_0\})$.
\end{rem}

\begin{rem}\label{uzirem}
Uzi Vishne has pointed out to me that one can distinguish a
random permutation \emph{which is not a cyclus} from a random cyclus
in with advantage $1$ at the price of increasing
the number of queries to $\nu(N)+1$ (where $\nu(N)$ is the number of prime divisors of $N$):
One simply verifies that for each prime factor $p$ of $N$,
$P^{N/p}(0)\neq 0$, whereas $P^N(0)=0$. This happens if, and only if, $P$ is a cyclus.
(Similar observations apply to Remarks \ref{rem1} and \ref{generalization}.)

Observe that in probability $1/N$, a random permutation is a cyclus
and therefore one cannot hope to obtain advantage greater than $1-1/N$,
so this improves the advantage from $1-d(N)/N$ to $1-1/N$ at the price
of $\nu(N)$ additional queries.
Clearly $\nu(N)\le\log_2 N$.
In fact, by the Hardy-Ramanujan Theorem, $\nu(N)$ is asymptotically close
to $\log\log N$ ``for almost all $N$'' (we will not give the precise
formulation here). Observe that when $N$ is a power of $2$ we
get here $\nu(N)=1$, so two queries are enough to distinguish
with advantage $1-1/N$.
\end{rem}

\Part{Fast forward random permutations}

This part introduces an efficient method to sample the cycle
structure of a random permutation, and its application to
the construction of fast forward random permutations.

\section{Ordered cycle structures}

\begin{definition}\label{OCS}
Assume that $\Om$ is a finite, well-ordered set, and $P\in S_\Om$.
Let $C_0,\dots,C_{k-1}$ be all (distinct) cycles of $P$, ordered such that
$\min C_i < \min C_j$ for each $i<j$.
Then the \emph{ordered cycle structure} of $P$, $\OCS(P)$,
is the sequence $(|C_0|,\dots,|C_{k-1}|)$.
\end{definition}
\begin{example}
If
$$P=\mx{0 & 1 & 2 & 3 & 4 & 5\\5 & 4 & 1 & 3 & 1 & 0}\mbox{,}$$
then the cycles of $P$ are $(0 5),(1 4 2),(3)$ in this
order, as the minimum elements of the cycles are $0,1,3$,
respectively. Thus, $\OCS(P)=(2,3,1)$.
\end{example}

Sampling the ordered cycle structure of a random permutation
in $P\in S_\Om$ (by choosing a random $P$,
finding the size of the cycle of $0$,
then the size of the cycle of the first element not in this cycle,
etc.) requires $O(|\Om|)$ steps,
which is infeasible when $\Om$ is a large space.
The following theorem allows us to sample this distribution
efficiently.

\begin{thm}\label{samplecyclen}
Let $\Om$ be a finite set of size $N$.
Consider the following two random processes:
\bi
\i[Process I:] Choose a random permutation $P\in S_\Om$,
and give $\OCS(P)$ as output.
\i[Process II:]
\be
\i Set $s_{-1}=0$.
\i For $i=0,\dots$ do the following:
\be
\i Choose a random number $s_i\in\{1+s_{i-1},\dots,N\}$.
\i If $s_i = N$, then exit the loop.
\ee
\i Output the sequence $(s_0,s_1-s_0,s_2-s_1\dots,s_i-s_{i-1})$.
\ee
\ei
Then these processes define the same distribution
on the space of all possible ordered cycle structures of permutations
$P\in S_\Om$.
\end{thm}
\begin{proof}
We prove the theorem
by induction on the size of $\Om$.
The theorem is evident when $|\Om|=1$.

For $|\Om|>1$, assume that $P$ is a random element of $S_\Om$, and let $\OCS(P)=(a_0,\dots)$.
By Lemma \ref{uniformcyclen},
$a_0$ distributes uniformly in $\{1,\dots,N\}$.
Using the notation of Definition \ref{OCS},
let $C_0$ be the cycle of $0$.
As $P$ distributes uniformly over $S_\Om$,
an easy counting argument shows that
the restriction of $P$ to the remaining elements,
$P\restriction\Om\sm C_0$ distributes uniformly over
$S_{\Om\sm C_0}$.
By the induction hypothesis,
the output $(b_0,b_1,\dots)$ of Process II
for $n=|\Om\sm C_0|$ distributes exactly as the output of
Process I on $P\restriction\Om\sm C_0$.
Thus, the sequence $(a_0,b_0,\dots)$ given by Process II distributes
the same as the sequence given by Process I.
\end{proof}

\begin{definition}
For ease of reference, we will call
Process II of Theorem \ref{samplecyclen}
the \emph{Choose Cycle Lengths (CCL)} process.
\end{definition}

Observe that the running time of the CCL process in the worse case
is $N$, which is too large (usually, a quantity which is
polynomial in $\log N$ is considered
small, and $\Omega(N^\epsilon)$ where $\epsilon>0$
is considered infeasible).
We can however define an algorithm which is probabilistically close
to the CCL process but runs in time $O(\log N)$.

Let $R_N$ denote the random variable counting the number of cycles in a
permutation in $S_N$.
It is well known \cite{SHL} that the expectation and variance
$R_N$ (and therefore the
running time of the CCL process)
are both $\log N + O(1)$.
By Chebyshev's Inequality,
\begin{eqnarray*}
\lefteqn{\Prob[R_N\ge (c+1)\log N] = \Prob[R_N-\log N\ge c\log N] =}\\
& = & \Prob[R_N-\log N\ge (c\sqrt{\log N})\sqrt{\log N}] \le\\
& \le & \f{1}{(c\sqrt{\log N})^2} = \f{1}{c^2\log N}
\end{eqnarray*}
for all constant $c>0$, which is $\Theta(1/\log N)$.
We say that a function $f(N)$ is \emph{negligible}
if it is $O(1/N^\epsilon)$ for some positive $\epsilon$.
The bound given by Chebyshev's Inequality
is not negligible. Fortunately we can improve it
significantly in our case. To this end, we need to have
a tight upper bound on the distributions of the
random variables $s_i$ defined by the CCL process.

\begin{prop}\label{s_idist}
Fix $l\in\{0,\dots,N-1\}$. Then
$$\Prob[s_l=k] < \f{\abs{\log(1-\f{k}{N})}^l}{l!N}$$
if $k\in\{l+1,\dots,N\}$ and is $0$ otherwise.
\end{prop}
\begin{proof}
Recall that for an increasing function $f:[0,k]\to\bbR$,
$\sum_{i=0}^{k-1} f(i) < \int_0^k f(x)dx$.

We prove the proposition  by induction on $l$.
For $l=0$ we have that $\Prob[s_0 = k] = 1/N$ as required.
Assume that our assertion is true for $l$, and prove it for
$l+1$ as follows.
\begin{eqnarray*}
\lefteqn{\Prob[s_{l+1}=k] =}\\
& = & \sum_{i=l+1}^{k-1}\Prob[s_l=i]\cdot\Prob[a_{l+1}=k-i | s_l=i] =
      \sum_{i=l+1}^{k-1}\Prob[s_l=i]\cdot\f{1}{N-i} <\\
& < & \int_0^k\f{\(-\log(1-\f{x}{N})\)^l}{l!N}\cdot\f{1}{N-x}dx
\end{eqnarray*}
Substituting $t=-\log (1-x/N)$, we have that the last integral is
equal to
$$\f{1}{l!N}\int_0^{-\log\(1-\f{k}{N}\)}t^l dt = \f{\(-\log\(1-\f{k}{N}\)\)^{l+1}}{(l+1)!N}.$$
\end{proof}

\begin{thm}\label{fastconverge}
Fix $l\in\{0,\dots,N-1\}$. Then for all $m$,
$$\Prob[s_l<m]<\f{m}{N}\cdot\f{\abs{\log\(1-\f{m}{N}\)}^l}{l!}.$$
\end{thm}
\begin{proof}
By Proposition \ref{s_idist},
\begin{eqnarray*}
\lefteqn{\Prob[s_l<m] <}\\
& < & \f{1}{l!N}\int_0^m\(-\log\(1-\f{x}{N}\)\)^l dx < \f{1}{l!N}\int_0^m\(-\log\(1-\f{m}{N}\)\)^l dx =\\
& = & m\cdot\f{\abs{\log\(1-\f{m}{N}\)}^l}{l!N}.
\end{eqnarray*}
\end{proof}

\begin{cor}\label{CCLrunningtime}
Assume that $c>e$.
The probability that the running time of the CCL process is larger
than $c\log N$ is
$O\(\sqrt{\log N}/N^{c(\log c -1)}\)$
and is therefore negligible.
In particular, if $c>e^2$ then this probability is $o(1/N^c)$.
\end{cor}
\begin{proof}
Use Theorem \ref{fastconverge} with $m=N-1$ and $l=c\log N$.
Then $1-m/N=1/N$.
Using Stirling's Formula,
\begin{equation}\label{eq}
\Prob[s_l<m] < \f{\abs{\log\f{1}{N}}^l}{l!} \approx \f{\log^l N}{\sqrt{\f{2\pi}{l}}\(\f{l}{e}\)^l}.
\end{equation}
Now, as $l=c\log N$,
$$\f{\log^l N}{\(\f{l}{e}\)^l} = \(\f{e\log N}{l}\)^l = \f{e^l}{c^l} = \f{N^c}{N^{c\log c}} = N^{c(1-\log c)},$$
therefore the right hand side of Equation \ref{eq} is equal to
$$\sqrt{\f{c\log N}{2\pi}}\cdot\f{1}{N^{c(\log c -1)}}.$$
This implies the assertions in the theorem.
\end{proof}

We can therefore define the following variant of the CCL process:
\begin{definition}[$l$-truncated CCL]
Fix a positive integer $l$ and run the CCL process $l-1$ steps.
If the process terminated after $k<l$ steps,
then output the sequence $(s_0,\dots,s_{k-1})$.
Otherwise
set $s_{l-1}=N$ and output $(s_0,\dots,s_{l-1})$.
\end{definition}

\begin{cor}\label{3.6logN}
Fix $l\ge 3.6\log N$. Then the output of the $l$-truncated
CCL cannot be distinguished from the output of the CCL process
with advantage greater than $o(1/N)$.
\end{cor}
\begin{proof}
This follows from Theorem \ref{CCLrunningtime}, once we observe (numerically)
that the solution to the equation $c(\log c-1)=1$ is $c=3.5911^+$.
\end{proof}

\section{Fast forward permutations}

\begin{definition}
Assume that $(a_0,a_1,\dots,a_{l-1})$ is a sequence of
positive integers such that $\sum_{k=0}^{l-1}a_k = N$, and
write $s_{-1}=0$, $s_i = \sum_{k=0}^i a_k$ for each $i = 0,\dots,l-1$.
The \emph{fast forward permutation coded by
$(a_0,a_1,\dots,a_{l-1})$} is the permutation
$\pi\in S_N$ such that for each $x\in\{0,\dots,N-1\}$,
$$\pi(x) = s_i + (x-s_i+1 \bmod a_{i+1})\quad\mbox{where }s_i\le x< s_{i+1}.$$
\end{definition}

\begin{example}
The fast forward permutation $\pi\in S_7$ coded
by $(1,2,4)$ is
$$\pi = (0)(12)(3456) = (12)(3456).$$
Here $s_0 = 1$, $s_1 = 3$, and $s_2 = 7$. Thus, e.g.,
as $s_1\le 4 < s_2$, we have that
$$\pi^5(4) = s_1+(4-s_1+5 \bmod a_2) = 3+(6 \bmod 4) = 5,$$
as can be verified directly.
\end{example}

A fast forward permutation coded by a sequence $(a_0,\dots,a_{l-1})$
is indeed fast forward, if we can either preprocess the corresponding
sequence $(s_0,\dots,s_{l-1})$ (this is done in time $O(l)$)
or have access to an oracle which can tell $s_i$ for each $i$ in time
$O(1)$.
\begin{prop}\label{ffcoded}
Assume that $\pi$ is the fast forward permutation coded by $(a_0,\dots,a_{l-1})$.
Assume further that
we have an $O(1)$ time access to the corresponding values $s_i$, $i\in\{0,\dots,l-1\}$.
Then for all $x\in\{0,\dots,N-1\}$ and all $m$,
the complexity of the computation of $\pi^m(x)$
is $O(\log l)$ (and in particular $O(\log N)$).
\end{prop}
\begin{proof}
As the values $s_i$ are increasing with $i$,
we can use binary search to find the $i$ such that $s_i\le x< s_{i+1}$
(this requires $O(\log l)$ accesses to the values $s_i$).
Then
$$\pi^m(x) = s_i + (x-s_i+m \bmod (s_{i+1}-s_i)).$$
\end{proof}
The proof of Proposition \ref{ffcoded} is written such that
we can see that the sequence $(a_0,\dots,a_{l-1})$ plays
no role in the evaluations of $\pi^m(x)$. This means that
all needed information is given in the sequence
$(s_0,\dots,s_{l-1})$. We chose the sequence
$(a_0,\dots,a_{l-1})$ rather than $(s_0,\dots,s_{l-1})$
as a ``code'' for the permutation only because this
way it seems more clear how the permutation $\pi$ is computed.

Consider the following oracles.
\bi
\i[$\PFF$:] Chooses a random permutation $P\in S_N$,
accepts queries of the form $(x,m)\in \{0,\dots,N-1\}\x \bbZ$,
and responds with $y=P^m(x)$ for each such query.
\i[$\mathcal{F}$:]
Runs the $l$-truncated
CCL process with $l=4\log N$ to obtain
a sequence $(a_0,\dots,a_{l-1})$.
(Let $\pi$ denote the fast forward permutation
coded by $(a_0,\dots,a_{l-1})$.)
This oracle
accepts queries of the form $(x,m)\in \{0,\dots,N-1\}\x \bbZ$,
and uses the oracle $\P$ (which fixes a random permutation $P$)
to respond with $y=P(\pi^m(P\inv(x)))$ for each such query.
\ei

\begin{thm}\label{FvsPFF}
\be
\i The space used by the oracle $\F$ is $O(\log N)$ words of size
$O(\log N)$ each.
\i The preprocess of $\F$ requires $O(\log N)$ steps.
\i For each query $(x,m)$, the running time of $\F$
is $O(\log\log N)$ plus twice the running time of $\P$.
\i Assume that $D$ is a distinguisher which makes any number of calls to
the oracles $\PFF$ or $\F$.
Then the advantage of $D$ is $o(1/N)$.
\ee
\end{thm}
\begin{proof}
(1) is evident. (2) follows from Proposition \ref{ffcoded},
and (3) follows from Corollary \ref{3.6logN}.
\end{proof}

This completes our solution to the Naor-Reingold Problem in the
(purely) random case.

\part{Pseudorandomness}\label{PRcase}

Intuitively speaking, pseudorandom objects are
ones which are easy to sample but difficult to
distinguish from (truly) random objects.
The assumption that we made on the oracle $\P$---namely,
that it chooses a random permutation in $S_N$---is
not realistic when $N$ is large. A more realistic assumption
is that the oracle chooses a \emph{pseudorandom} element
of $S_N$. More concretely, the oracle $\P$ accepts a \emph{key}
$k$ as input, and uses it to define a permutation $P_k$
in the sense that each time the oracle is asked to compute
$P_k(x)$ (or $P_k\inv(x)$), the oracle computes it without
the need to explicitly build the complete permutation $P_k$.
($\P$ can be thought of as a key dependent block cipher.)
The reader is referred to \cite{NaRe} for the formal definitions.
Naor and Reingold \cite{NaRe} actually stated their problem
in the pseudorandom case.
We will translate our main results into the pseudorandom case.

\section{Translation of results from Part 1}

Let $\C'$ be a pseudorandom cyclus oracle.
This means that
for any distinguisher $D$ which makes
a small number $m$ of queries,
the advantage $a=|\Prob[D(\C')=1]-\Prob[D(\C)=1]|$
is small.
\begin{thm}\label{PRcycvsperm}
For any distinguisher $D$ which makes $m<N$ queries to
$\C'$ or $\P$,
$$|\Prob[D(\C')=1]-\Prob[D(\P)=1]|\le a+\f{m}{N},$$
where $a=|\Prob[D(\C')=1]-\Prob[D(\C)=1]|$.
\end{thm}
\begin{proof}
By the Triangle Inequality and Theorem \ref{cycvsperm},
\begin{eqnarray*}
\lefteqn{|\Prob[D(\C')=1]-\Prob[D(\P)=1]| \le}\\
& \le & |\Prob[D(\C')=1]-\Prob[D(\C)=1]|+|\Prob[D(\C)=1]-\Prob[D(\P)=1]|\le\\
& \le & a+\f{m}{n}.
\end{eqnarray*}
\end{proof}

\begin{thm}\label{PRoptimal}
Consider the $m$-step strategy ($m<N$) for a distinguisher $D$
which was defined in Theorem \ref{optimal} (an arbitrary strategy which
generates nonrepeating sequences.)
Then
$$|\Prob[D(\C')=1]-\Prob[D(\P)=1]|=\f{m}{N}.$$
 Consequently, for all $\epsilon>0$ there exists a strategy $D$ to distinguish
$\C'$ from $\P$ with advantage $\max\{a-\epsilon,m/N\}$,
where $a$ is the supremum of all possible advantages of
an $m$-step distinguisher to distinguish
$\C'$ from $\C$.
\end{thm}

\begin{proof}
The proof of Theorem \ref{optimal} only uses the fact that $\P$
chooses a random permutation and $\C$ chooses a cyclus.
The fact that the cyclus $\C$ is random is not used.
This implies the first claim in our theorem.

To prove the second part of the theorem,
fix any $\epsilon>0$.
If $a-\epsilon\le m/N$, we choose the strategy $D$ and we are done.
Otherwise $m/N<a-\epsilon$.
As $a-\epsilon<a$,
there exists an $m$-step strategy
$D'$ to distinguish $\C'$ from $\C$ with advantage at least
$a-\epsilon$, so we can choose the strategy $D'$.
\end{proof}

We now translate the main result in the fast forward model
to the pseudorandom case.
\begin{thm}\label{PRcheckcyclus}
$\C'$ can be distinguished from
$\P$  with advantage $1-d(N)/N$,
using a single query.
\end{thm}
\begin{proof}
Again, the only property of $\C$ we used in the proof of Theorem
\ref{checkcyclus} is its choosing a cyclus, which is also
true for $\C'$.
\end{proof}

\section{Translation of results from Part 2}

In order to shift to the pseudorandom case
in our construction of a fast forward permutation,
we need to have some pseudorandom number generator to generate
the random choices of the $s_i$'s in the CCL process.
If we have no such generator available, we can use
the oracle $\P$ itself: In addition to the key $k$ used
to generate $P_k$, we need another key $\tilde k$.
The pseudorandom numbers $s_i$ in the CCL process can
then be derived from the values $P_{\tilde k}(0)$,
$P_{\tilde k}(1)$, $P_{\tilde k}(2),\dots$
(This is the standard \emph{counter mode}
\cite{SCH}). We now give an example how this can
be done.

Consider the following oracles.
\bi
\i[$\RND$:] Accepts positive integers $x,k<N$ and returns
a sequence $(r_0,\dots,r_{k-1})$ of random numbers in the range $\{0,\dots,x-1\}$.
\i[$\RND_1$:] Accepts positive integers $x,k<N$, calls
$\RND$ with $N$ and $2k$ to get a sequence $(x_0,\dots,x_{2k-1})$,
and returns $(r_0,\dots,r_{k-1})$ where $r_i = (x_{2i}+N\cdot x_{2i+1}) \bmod x$
for all $i=0,\dots,k-1$.
\i[$\RND_2$:] Accepts positive integers $x,k,p_0<N$, calls
$\P$ $2k$ times to obtain the sequence $(x_0=P(p_0),\dots,x_{2k-1}=P(p_0+2k-1 \bmod N))$,
and returns $(r_0,\dots,r_{k-1})$ where $r_i = (x_{2i}+N\cdot x_{2i+1}) \bmod x$
for all $i=0,\dots,k-1$.
\ei

\begin{thm}\label{RNDvsRND}
Fix positive integers $x,k<N$. Then:
\be
\i If $k=c\log N$, then
$\RND$ and $\RND_1$ called with $x$ and $k$ cannot be distinguished with advantage greater than
$c\log N/N$.
\i $\RND_1$ and $\RND_2$ called with $x$ and $k$ cannot be distinguished with advantage greater than
$2k^2/N$.
\ee
\end{thm}
\begin{proof}
(1) Assume that $a$ and $b$ are random numbers in the range $\{0,\dots,N-1\}$.
Then $c=a+bN$ is random in the range $\{0,\dots,N^2-1\}$.
Let $x\in\{0,\dots,N-1\}$. With probability at least $1/N$,
$c<\lfloor N^2/x\rfloor\cdot x$ and therefore $c\bmod x$ is
random in the range $\{0,\dots,x-1\}$.
The probability that this happens $c\log N$ times
is therefore at least
$(1-1/N)^{c\log N}\approx e^{-c\log N/N}>1-c\log N/N$.

(2) This follows from the well known result that a random permutation is
a pseudorandom function. Briefly (see \cite{birthday} for more details),
consider any sequence of $2k$ random numbers in the range $\{0,\dots,N-1\}$.
The probability that all these numbers are distinct is greater than $1-(2k)^2/2N=1-2k^2/N$,
and in this case this sequence forms a random partial permutation.
\end{proof}

Consider now the modification $\F'$ of the oracle $\F$ which calls $\P$
with two independent keys $k$ and $\tilde k$, one for the evaluations
$P_k(\pi^m(P_k\inv(x)))$ and the other for the values
$P_{\tilde k}(0),P_{\tilde k}(1),\dots$ to be used by $\RND_2$
in order to generate the sequence of pseudorandom numbers
required by the $l$-truncated CCL process (the input argument $p_0$ to
$\RND_2$ is used to avoid sampling the same entry of $P_{\tilde k}$ twice).

\begin{thm}
$\F'$ and $\F$ cannot be distinguished with advantage greater than
$O(\log^2 N/N)$.
\end{thm}
\begin{proof}
This follows from the Triangle Inequality and the earlier results
\ref{FvsPFF}, \ref{RNDvsRND}(1), and \ref{RNDvsRND}(1) with $k=4\log N$.
\end{proof}

Here too, using a pseudorandom permutation oracle $\P'$ instead of a random one
in the definition of $\F'$ cannot increase the advantage by more than
$a$ where $a$ is the maximal advantage obtainable in distinguishing
$\P$ from $\P'$.

\section{Final remarks and open problems}

Another problem is mentioned in the original paper of Naor and Reingold
\cite{NaRe} and remains open, namely, whether one can construct a family
of fast forward pseudorandom \emph{functions} with graph structure
distribution similar to that of pseudorandom functions.

The natural analogue of our construction for the case of
pseudorandom permutations would not work for pseudorandom functions,
simply because the ``graph structure'' of a pseudorandom function carries
too much information. For example, there are $O(N)$ points with no
preimage. This was not the case with permutations, where the
structure is determined by the logarithmic number of its cycles
and their length. Another approach will be needed in order to solve this
problem.

Our study raises some other interesting open problems,
the most interesting of which seems to be the following.
Consider the $l$-truncated CCL process with $l=\log N$,
which uses an oracle $\RND_3$ similar to $\RND_2$ as its random
number generator with the difference that it makes
only $k$ calls to $\P$ to generate $(x_0=P(p_0),\dots,x_{k-1}=P(p_0+k-1 \bmod N))$,
and uses $r_i = x_i \bmod x$
instead of the original definition.
(So we use $\log N$ values of $P$ instead of $8\log N$ in the current construction.)
The problem is to prove or disprove the following.
\begin{conj}
$\F'$ with the parameters just described cannot be distinguished from $\PFF$
with a non-negligible advantage.
\end{conj}

\section{Acknowledgments}
I thank Kent E.\ Morrison for reference \cite{SHL}, and
Uzi Vishne for reading the paper and suggesting (the first paragraph of)
Remark \ref{uzirem}.
A special thanks is owed to Moni Naor for encouraging me to publish
these results, and to the referee for suggesting important improvements in
the presentation of the paper.


\begin{thebibliography}{00}


\bibitem{NaRe}
Moni Naor and Omer Reingold,
\emph{Constructing Pseudo-Random Permutations with a Prescribed Structure},
Journal of Cryptology \textbf{15} (2002),
97--102.

\bibitem{SCH}
Bruce Schneier,
\emph{Applied Cryptography},
John Wiley and Sons,
1996.


\bibitem{SHL}
L.\ A.\ Shepp and S.\ P.\ Lloyd,
\emph{Ordered cycle lengths in a random permutation},
Transactions of the American Mathematical Society \textbf{121} (1966), 340--357.

\bibitem{birthday}
B.\ Tsaban,
\emph{Bernoulli numbers and the probability of a birthday surprise},
Discrete Applied Mathematics \textbf{127} (2003), 657--663.


\end{thebibliography}
\end{document}